\documentstyle[11pt,newpasp,twoside,epsf]{article}
\def\edcomment#1{\iffalse\marginpar{\raggedright\sl#1\/}\else\relax\fi}
\reversemarginpar

\begin{document}
\title{Abundances in stars with exoplanets}
\author{Garik Israelian}
\affil{Instituto de Astrof{\'\i}sica de Canarias, E-38205 La Laguna,
Tenerife, Spain \\
e-mail gil@iac.es
}

\begin{abstract}
Extensive spectroscopic studies of stars with and without planets have concluded that stars hosting 
planets are significantly more metal-rich than those without planets. More subtle trends of different 
chemical elements begin to appear as the number of detected extrasolar planetary systems continues 
to grow. I review our current knowledge concerning the observed abundance trends of various
chemical elements in stars with exoplanets and their possible implications.

\end{abstract}

\section{Introduction}

Beginning with the discovery by Mayor \& Queloz (1995) of a giant planet racing around 51 Pegasi,
the number of planets orbiting solar-type stars has now reached 115. Most of the planets have been
discovered by the Geneva and California \& Carnegie groups led by Michel Mayor (Geneva),
Geoff Marcy (California) and Paul Butler (Carnegie).  This sample size 
is now sufficient to search for various trends linking the properties of exoplanets 
and those of their parent stars. It has been suggested that one of the key factors relevant to 
the mechanisms of planetary system formation is the amount of metals available 
in proto-planetary discs. 

The chemical abundance studies
of planet hosts are based on high signal-to-noise (S/N) and high resolution spectra.  
Many targets have been observed by more than one group, allowing useful cross checks of their
analyses and spectra. Most chemical studies of the planet hosts used iron as the reference
element (Gonzalez 1997, Laws et al. 2003, Murray \& Chaboyer 2002,  Santos, Israelian \& 
Mayor 2001, 2003a, Santos et al. 2003b) and only a few studies have discussed the abundance trends of 
other metals (Gonzalez \& Laws 2000, Sadakane et al. 2003, Gonzalez et al. 2001, Santos, Israelian \& Mayor 2000, 
Bodaghee et al. 2003). The authors in most of these studies have been constrained to compare the 
results for the planet host sample with other studies in the literature. However,
in some articles such a comparison was not provided, leaving room for any 
kind of speculation regarding the source of the abundance anomalies. Different authors used
different sets of lines, atmospheric parameters, data, etc. These are all potential sources of 
systematic error. To overcome this problem, Santos et al. (2001) prepared a sample 
of stars without known planets. To ensure a high degree of consistency between the two samples, 
these stars were analyzed and observed in the same way as the planet hosts. Further spectroscopic
analysis by Santos et al. (2003a, 2003b), Israelian et al. (2003a) and Bodaghee et al. (2003) were
based on this same comparison sample. 

Chemical abundance studies of planet host stars have revealed that their metallicities are higher on average
than those typically found among solar-type disk stars without known planets (Gonzalez 1997, 
Laws et al.\,2003, Santos et al.\,2001, 2003a, 2003b). In Fig.\,1 a comparison between the metallicity 
distributions for a volume-limited comparison sample of
stars and for the 87 planet hosts from Santos et al. (2003a) is presented.  The stars with planets are metal-rich
compared with the comparison sample stars by, on average, 0.25 dex. This suggests that the 
metallicity and the presence of giant planets are linked.  The metallicity excess could result 
from the accretion of planets and/or planetesimals onto the star (Gonzalez 1997). 
Opposing this view, Santos et al. (2000, 2001) proposed that the source of the high metallicity
is primordial and the observed abundance trends represent those from the proto-planetary and proto-stellar
molecular cloud out of which the star and the planets formed. This idea would support the classical
CIA (core-instability accretion) model (Pollack et al. 1996) where some 10-15 M$_{\earth}$ masses of 
planetesimals condense into a rocky core. The initial metallicity of the parental cloud is a key
parameter in this scheme. Results from Bodaghee et al. (2003) clearly demonstrate
that the excess in metallicity observed for planet host stars is widespread and not unique to iron. 
Abundances of different elements may provide clue for checking various planet formation,  and 
even planet migration, hypothesis. The self-enrichment scenario (Gonzalez 1997) which has been
proposed to explain the [Fe/H] excess in host stars should lead to a relative overabundance of
refractory elements (iron-group, $\alpha$-elements etc.) compared to volatiles (C, N, O, S, Zn). 
Volatiles are known to condense into solid grains at relatively low temperatures, and are expected 
to behave differently compared to the refractories which condense at high temperature. 
If the star accreted a considerable amount of planetary material, then high temperatures near the
star would favor the addition of refractory elements over volatiles (which are locked in giant
planets) and a trend in abundance versus condensation temperatures may appear (Smith et al. 2001).

\begin{figure}
\plotone{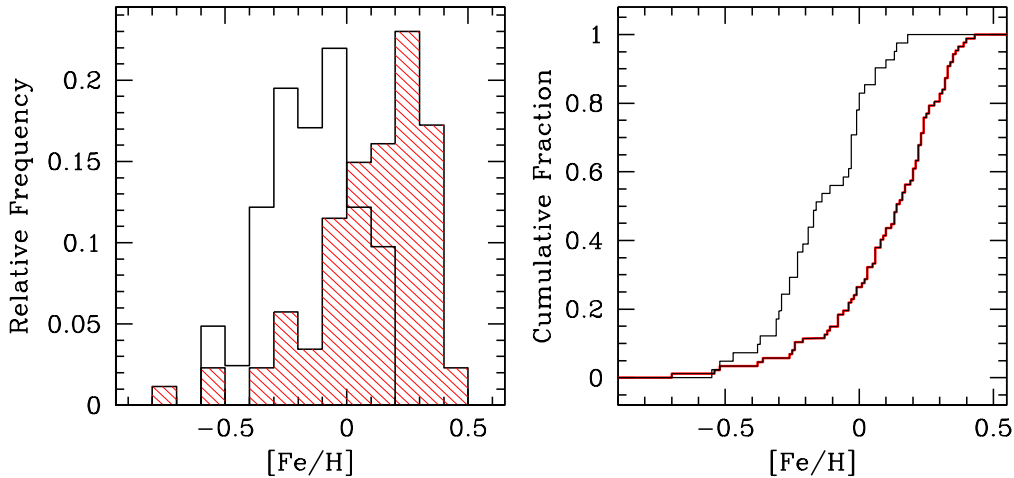}
\caption{Metallicity distribution for stars with planets (hatched histogram) compared with the
same distribution for the field stars without planets (empty histogram).
A Kolmogorov--Smirnov test shows the probability for the two populations being  part
of the same sample to be 10$^{-8}$. From Santos et al. (2003a).}
\end{figure}

\section{Abundances of light elements}

The light elements Li and Be are important tracers of the internal structure and
pre-main sequence evolution of solar type stars since they provide 
information regarding the redistribution and mixing of matter within a 
star. Studies of Be and Li complement each other as Li is depleted at much lower 
temperatures than Be. By measuring Li and Be in stars hosting planets we can obtain crucial 
information about the mixing, diffusion and angular momentum history of the stars. 
Accretion of planets and planetesimals, stellar activity and tidal interactions in
star-planets systems may largely modify the surface abundances of the light elements.

\subsection{Lithium}

Gonzalez \& Laws (2000) presented a direct comparison of Li abundances among 
planet-harbouring stars with field stars without planets and proposed that the 
former have less Li. However, in a critical analysis of this problem 
Ryan (2000) concludes that planet hosts and field stars have similar Li abundances. 
Given the larger number of planet-harbouring stars, we re-investigated the Li problem
(Israelian et al. 2003a) and looked for various statistical trends. 
When the Li abundances of planet host stars are compared with the 157 field stars
in the sample of Chen et al.\, (2001), we find that the Li abundance
distributions in the two samples are different (Fig.\,2). 
There is a possible excess of Li depletion in planet hosts having effective
temperatures in the range 5600--5850 K, whereas we find no significant differences 
for stars with temperatures in the range 5850--6350 K (Fig. 3). One may ask, why is 
the difference only seen in stars with effective temperatures in the range 5600--5850 K ?
Given the depth of the surface convection zone, we expect that any effect on the Li
abundance will be more apparent in solar-type stars. Lower mass stars have deeper 
convective zones and destroy Li  more efficiently, so we can often only set upper limits to the
abundance. However, the convective layers of stars more massive than the Sun do not
reach the lithium burning layer and therefore these stars generally preserve 
a large fraction of their original Li. The relatively small dispersion of Li 
abundances in these hotter stars is clearly seen in Fig.\,3. 
Therefore it seems that solar-type stars are the best targets 
for investigating any possible (and maybe marginal) effects of planets on the 
evolution of the stellar atmospheric abundance of Li. 

There are at least two possible hypothesis for the lower Li abundance in
planet-hosting stars. It is possible that proto-planetary disks lock a lot of
angular momentum and therefore create some rotational braking in the host 
stars during their pre main-sequence evolution. The lithium is efficiently 
destroyed during this process due to an increased mixing.  
The extra Li depletion can also be  associated with a planet migration 
mechanism at early times in the evolution of the star when the superficial convective 
layers may have been rotationally decoupled from the interior. Efficient depletion may 
be caused by a strong mixing due to the  migration-triggered  tidal forces, which 
create a shear instability. The mass of the decoupled convection zone in these 
stars is comparable to the masses of the known exoplanets; therefore, the migration 
of one or more giant planets could indeed produce an observable effect. The planetary 
migration may also trigger the accretion of planetesimals, inducing metallicity
enhancement. Some fresh Li could also be added in the convective zone. However, if this 
process takes place in the early evolution of the star, the freshly added Li will 
be destroyed.

\begin{figure}
\plotfiddle{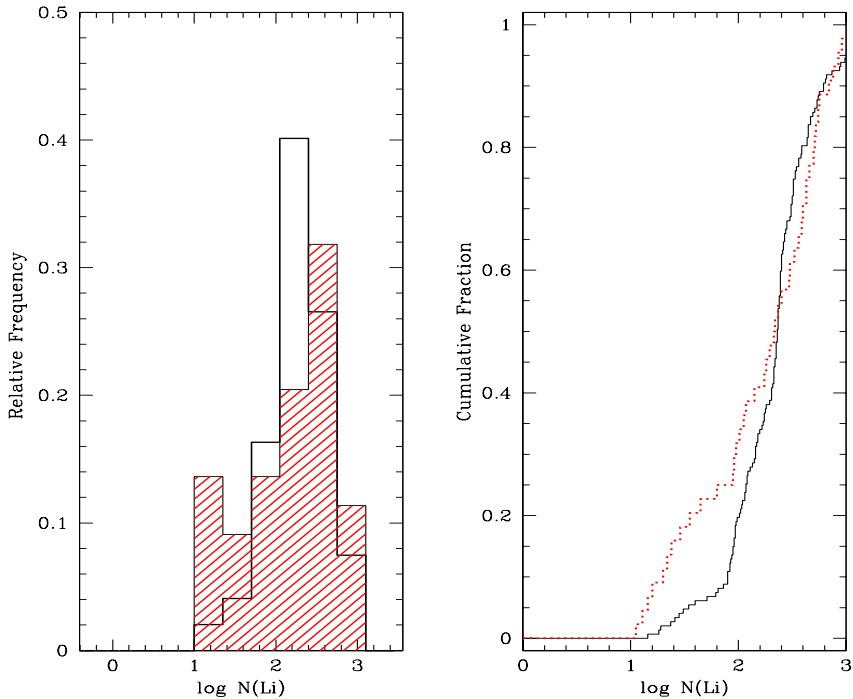}{8.5cm}{360}{60}{50}{-180}{-90}
\caption{Lithium distribution for stars with planets (hatched histogram) compared
with the same distribution for the field stars from Chen et al.\ 2001. (empty histogram).
A Kolmogorov--Smirnov test shows the probability for the two populations being a part of
the same sample is 0.2. From Israelian et al. (2003a).}
\end{figure}

Observations show a possible lack of short-period massive planets with $M >$ 4 $M_{\rm J}$ 
orbiting around Li depleted stars with 1 $< \log \epsilon({\rm Li}) < $1.6  (Israelian et al. 2003a). 
When considering stars with $\log \epsilon({\rm Li}$) between 1.6 and 3, we find that about half 
have planets with  $M>$ 4 $M_{\rm J}$. There is a clear link between  Mass--Li and Period--Li relationships 
most likely associated with the already proposed correlation between mass and period of planetary
companions  (Zucker \& Mazeh 2002; Udry et al.\ 2002).

\subsection{The $^6$Li--test}

A unique opportunity for testing the planet and/or planetesimal accretion
scenario is offered by a $^6$Li--test proposed by Israelian et al. (2001).
This approach is based on looking for an element that should not appear
in the atmosphere of a normal solar-type star, but would be present in a star that
has accreted planetary matter. 
Nuclear reactions destroy the $^6$Li and $^7$Li isotopes in
stellar interiors at temperatures of $2\times10^6$ ($^6$Li) and $2.5\times10^6$ K
($^7$Li). Furthermore, convection cleans the upper atmosphere of Li nuclei by transporting 
them to deeper and hotter layers where they are rapidly destroyed. Young solar-type stars 
are entirely convective and most of the primordial Li nuclei are burned in their interiors
in a mere few million years. However, many solar-type
stars preserve a large fraction of their initial atmospheric $^7$Li nuclei.
According to standard models (Forestini 1994), at a given metallicity there is a mass
range where $^6$Li, but not $^7$Li, is destroyed. These models predict
that no $^6$Li can survive pre-MS mixing in metal-rich solar-type stars.
The detection of $^6$Li in HD\,82943 (Israelian et al. 2001, 2003b) is convincing
observational evidence that stars may accrete planetary material, 
or even entire planets, during their main sequence evolution. Other explanations of this 
phenomenon  such as stellar flares or surface spots have been ruled out (Israelian et al. 2001).
Sandquist et al. (2002) have recently proposed that $^6$Li can be used to distinguish between 
different giant planet formation theories.

However, analysis of $^6$Li is difficult. First of all, it is a weak component of a 
blend with much stronger doublet of $^7$Li having an isotopic separation of 0.16 \AA. Blending 
of the Li line with other weak absorptions and the placement of continuum pose serious problems
in metal-rich solar-type stars. Spectra with S/N $\sim$ 1000 and a resolving power of at 
least $\lambda / \Delta \lambda \sim$ 100.000  are required to tackle
these problems. In metal-rich stars the identification of any weak blends in the
region of the Li absorption becomes crucial. For example,  Reddy et al.\,(2002) 
claimed that a previously noticed weak absorption in the solar spectrum at 6708.025 \AA\  
belongs to \ion{Ti}{i}. With this assumption their study of the Li feature in 
HD\,82943 did not confirm the presence of $^6$Li. However, our recent analysis 
(Israelian et al. 2003b) does not support the identification of a weak absorption 
feature at 6708.025 \AA\ with the low excitation \ion{Ti}{i} line. We have suggested 
that the unidentified absorption is most
probably produced by a high excitation \ion{Si}{i} line.  
The presence of $^6$Li in HD\,82943 was confirmed 
with the updated value for the isotopic ratio $f(^6$Li) = $0.05\pm0.02$ by taking 
the \ion{Si}{i} line into account in the reanalysis of the $^6$Li/$^7$Li
and using new VLT/UVES spectra with S/N $\sim$ 1000.

Slow accretion of  planetesimals was invoked in order to explain the [Fe/H]
distribution in planet-harbouring stars. Recently, Murray \& Chaboyer (2002)
concluded that an average of 6.5\,$M_{\earth}$ of iron must be added to the 
planet-harbouring stars in order to explain the mass--metallicity and age--metallicity 
relations. Accretion of 6.5\,$M_{\earth}$ of planetesimals of iron during early MS evolution will
strongly modify $^7$Li abundances in these stars. Moreover, given the depth of the 
convection zone in stars with T$_{eff}$ $>$ 5900 K, a large amount of the added $^6$Li may
avoid  destruction via mixing. Accretion of a chondritic matter with 
6.5\,$M_{\earth}$ of iron by a star with $T_{eff}$=6100 K and with a convection zone mass 
10$^{-3}$\,M$_{\sun}$ will rise its $^7$Li abundance from $\log \epsilon({\rm Li})$=2.7 to 3.2 
while the isotopic ratio will become $f(^6$Li) = 0.06. This will create a detectable $^6$Li 
absorption feature with an equivalent width (EW) $\sim$\,4 m\AA. This feature can be measured 
even if it is blended with the line at 6708.025 \AA\ because the latter is expected to appear 
with an EW $\la$ 2 m\AA\ in these type of stars (no matter which kind of element/line 
is responsible for this absorption).

\subsection{Beryllium}

The first studies of Be in planet hosts were those by Garc\'\i a L\'opez \& P\'erez de Taoro (1998) 
and Deliyannis et al.\,(2000). However, these authors did not arrive to any firm conclusion because of
the lack of a comparison sample of stars and the low number of planet hosts in their studies. 
Santos et al.\,(2002) derived beryllium abundances for a sample of 29 planet host and 6 ``single'' stars 
aimed at studying in detail the effects of the presence of planets on the structure and evolution of the 
associated stars. Their preliminary results suggest that theoretical models may have to be revised for
stars with T$_{eff} <$ 5500 K. Santos et al.\,(2002) found several Be depleted stars at 5200 K which current 
models cannot explain. A comparison between planet-hosting stars and "single" stars (although very few in 
their analysis) shows no clear difference between either population. Their preliminary result 
supports a ``primordial'' origin for the metallicity excess observed in the planet hosts. 
The analysis of a comparison sample will certainly help to further constrain the models.

\begin{figure}
\plotfiddle{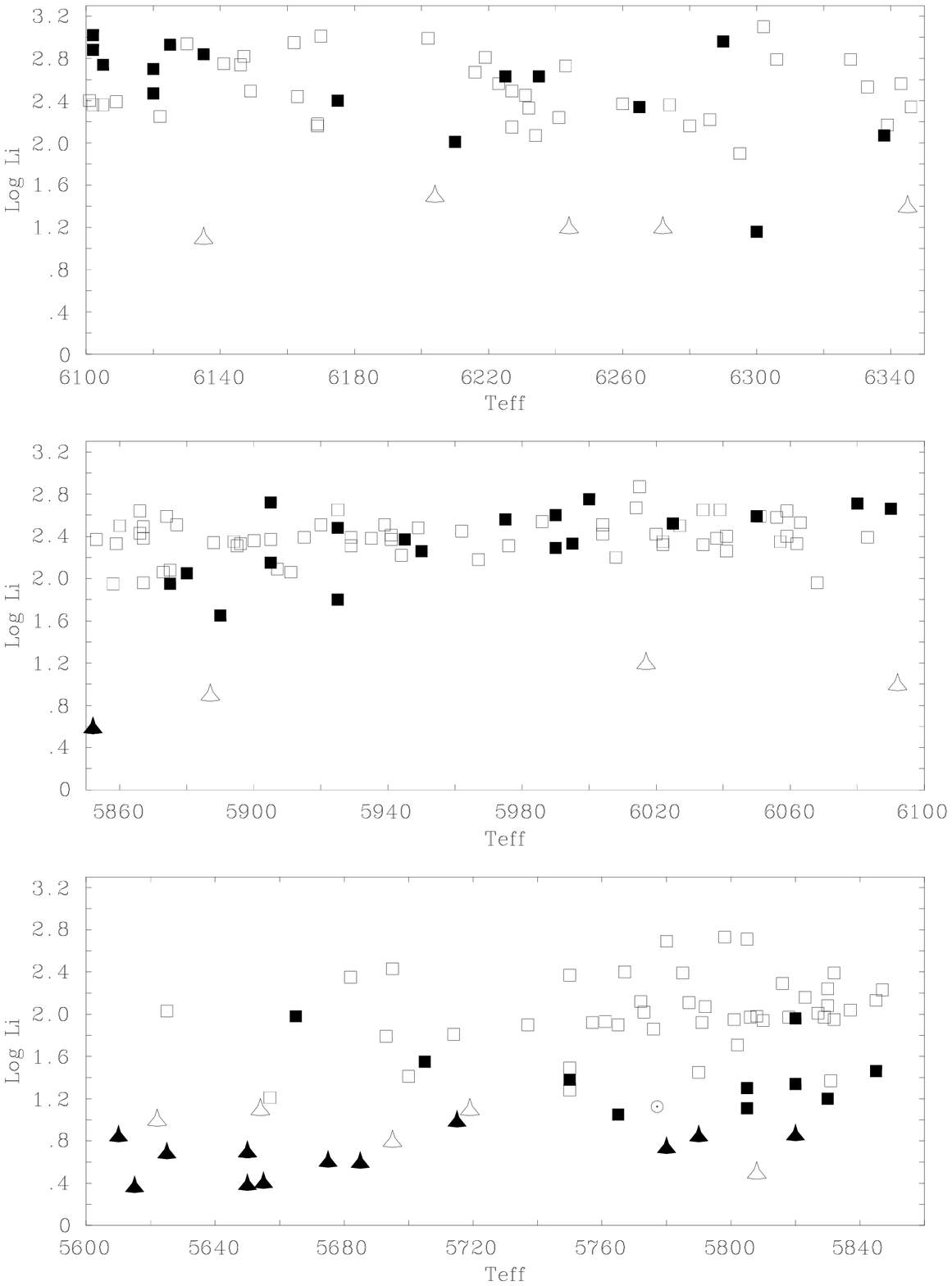}{9.0cm}{360}{60}{50}{-180}{-55}
\caption{Lithium versus effective temperature for stars with planets
(filled squares) and the comparison sample of Chen et al.\ (2001) (empty squares). Upper
limits are filled (planet hosts) and empty (comparison sample) triangles.
The position of the Sun is indicated. From Israelian et al. (2003a).}
\end{figure}

\begin{figure}
\plottwo{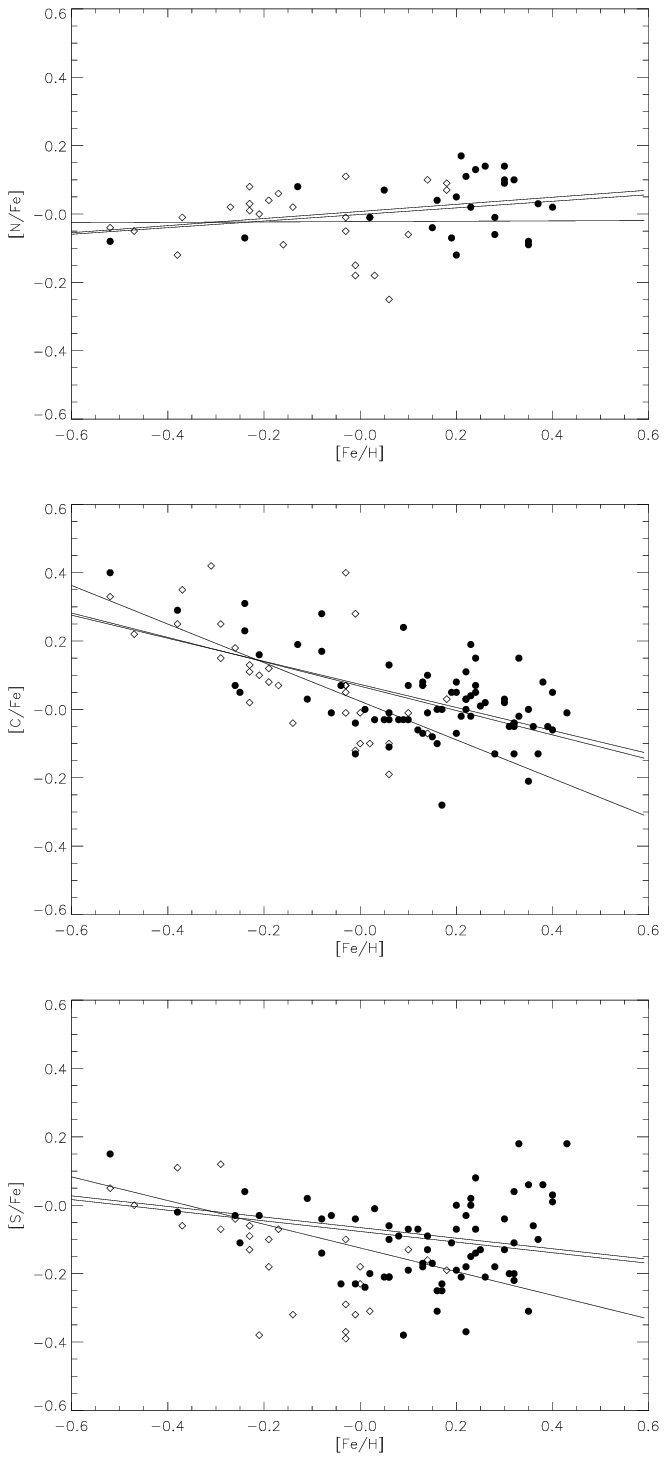}{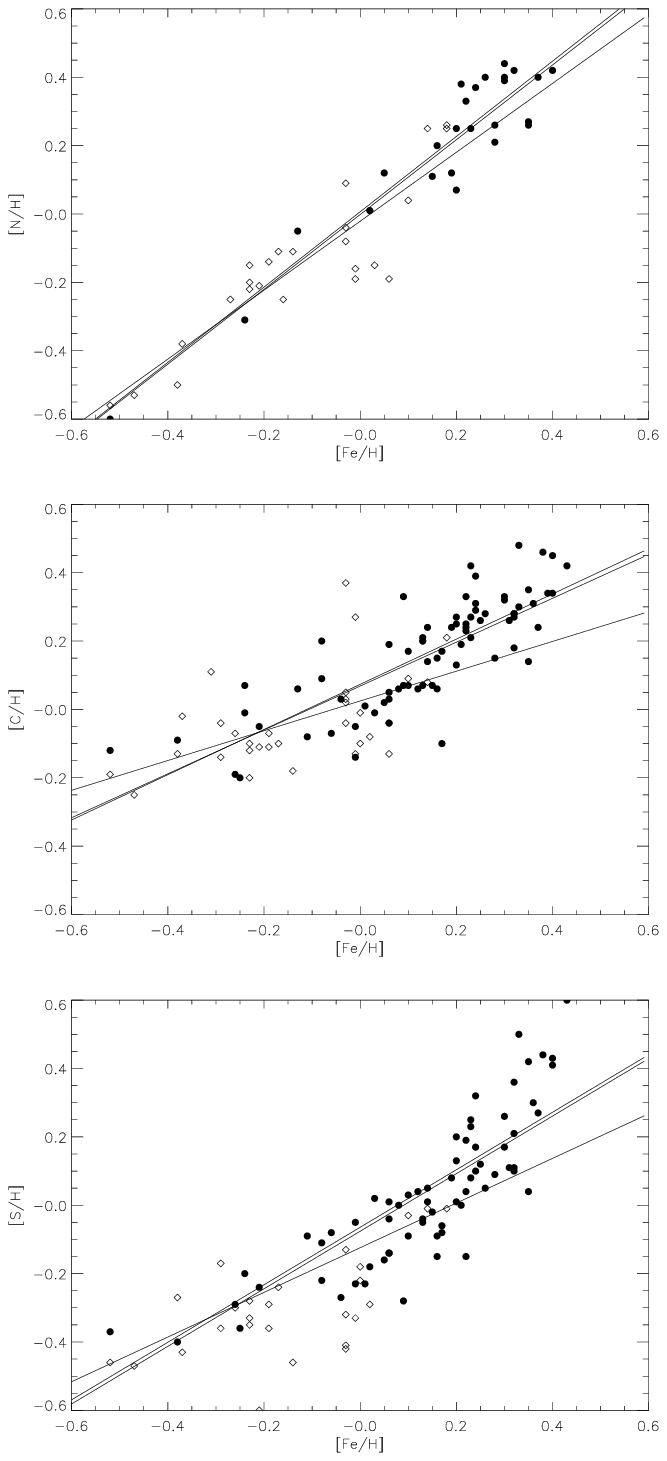}
\caption{[X/Fe] vs. [Fe/H] plots {\em (left)} and [X/H] vs. [Fe/H] plots {\em (right)} 
for N, C and S. Filled circles represent planet host stars, while open symbols denote comparison
sample stars. Linear least-square fits to the planet hosts {\em(dotted lines)}, to
the comparison sample stars {\em(dashed lines)} and to both samples together {\em(solid
lines)} are represented. Only the stars with $T_{\em eff}>5200\,K$ were considered in 
the analysis of S. From Ecuvillon et al. (2003).}
\end{figure}

\section{Abundances of metals}

\subsection{Volatiles}

Santos et al.\,(2000) and Gonzalez \& Laws\,(2000) have noticed that planet hosts tend to show
subsolar [C/Fe] with increasing [Fe/H]. Gonzalez et al.\,(2001) were less certain about these
findings, while Takeda et al.\,(2001) and Sadakane et al.\,(2002) found no clear evidence of
the constant [C/Fe] in the metallicity range $-0.5 \la$ [Fe/H] $\la$ 0.4.
However, all these authors used field stars from the literature to build up their small and
inhomogeneous comparison samples. For example, Sadakane et al. (2002) used for C and O 
the results of the analysis by Gustaffson et al.\,(1999), Edvardsson et al.\,(1993),
Chen et al.\,(2000) and Feltzing \& Gustaffson\,(1998). They concluded that [C/Fe]
and [O/Fe] ratios in planet hosts closely follow the trends observed in field stars with
[C/Fe]=[O/Fe]=0 at [Fe/H]$>$0. Although N abundances were derived by Gonzalez \& Laws\,(2000), 
Gonzalez et al.\,(2001) and Sadakane et al.\,(2002), none of these authors discussed the 
trends of [N/Fe] and [N/H].

Recently, Ecuvillon et al.\,(2003) used VLT/UVES spectra in order to derive the N abundance 
in a large number of planet hosting stars and comparison sample stars from Santos et al.\,(2001). 
The  near-UV NH band at 3340-3380\,\AA\ was employed in their analysis. In addition, they
studied C and S abundances from several optical lines. Their results indicate a clear 
difference in [N/H], [C/H] and [S/H] distributions for both samples. They found 
that [S/Fe] and [N/Fe] are flat at [Fe/H]$>$0. The final abundance ratios of
C, N and S for both samples as functions of [Fe/H] are displayed in Fig.\,4. 
The trend of [C/Fe] is similar to the [O/Fe] trend; monotonically decreasing at higher 
[Fe/H], as discussed by Bensby et al. (2003) and Feltzing \& Gustaffson\,(1998). 
Finally we note a small increase in [S/Fe] at [Fe/H] $> $0.2 (see Fig.\,4) also seen in
the Fig.\,10 of Sadakane et al.\,(2002). We cannot rule out the possibility that the [S/H] ratios 
of planet-bearing stars are not an iron-rich extension to the [S/H] trends of the comparison sample. 
The behaviour of [X/H] versus [Fe/H] for some volatiles coupled with the presence of giant
planets merits further investigation.

\subsection{Refractories}

The first abundance studies of several refractory elements  (Gonzalez 1997, Gonzalez \& Laws 2000,
Gonzalez et al. 2001, Santos et al. 2000) revealed a few possible anomalies. 
Gonzalez et al. (2001) claimed that the stars with planets appear to have smaller [Na/Fe], [Mg/Fe] 
and [Al/Fe] values than field dwarfs of the same [Fe/H]. These authors did not find any
significant differences for the refractories Si, Ca and Ti. However, the abundance trends in the few planet 
hosts discussed by Takeda et al. (2001) and Sadakane et al.\,(2002) did not show anything peculiar. 
On the other hand, anomalies were found by Sadakane et al.\,(2003) who detected a few planet-bearing stars 
with an interesting abundance pattern in which the volatile elements C and O are underabundant with respect 
to refractories Si and Ti. 
 
A uniform and unbiased comparison of abundances of some $\alpha$- (Si, Ca, Ti) and Fe-group
(Sc, V, Cr, Mn, Co, Ni) elements in 77 planet host and 42 comparison sample stars without planets 
was carried out by Bodaghee et al. (2003). These authors concluded that the abundance trends for 
the planet hosts are almost identical to those in the field. Slight differences were found for 
V, Mn and, to a lesser extent, Co and Ti (Fig.\,5). Although the abundance scatter 
for most of the elements was found to be small, a few elements showed considerable dependence of 
the derived abundances on the effective temperature. The largest effect was found for Ti, Co 
and V, where the difference between K and F-dwarfs has reached 0.2-0.3 dex. These trends might be
related to NLTE effects. 

In general, the abundance distributions of planet host stars are high [Fe/H] extensions to the
curves traced by the field dwarfs without planets. No significant differences are found in the regions
of overlap. However, although some differences for certain elements are subtle (and may even be 
negligible), they are certainly intriguing enough to merit additional studies.

\section{Implications}

\subsection{Chemical Evolution of the Galaxy}

One of the byproducts of chemical abundance studies in planet-hosting stars is the possibility of 
learning about Galactic chemical evolution trends at high metallicities. The number of 
detailed abundance studies at [Fe/H]$>$0 is very limited and exoplanet hosts can help 
to explore this regime. Some of the trends obtained in these studies may be linked with 
the presence of giant planets. According to Santos et al.\,(2003a), more than 25\,\% of stars
with [Fe/H] $>$0.3 host planets.  The possibility that all metal-rich stars host planetary 
systems cannot be ruled out. Thus, it is almost impossible to compare stars with
and without planets in the high [Fe/H] tail of the distribution. 
The relative frequency of stars with planets increases with [Fe/H], but there is a
sharp cutoff once the metallicity reaches about 0.4 dex (Fig.\,1). It is hard to believe that
Nature could somehow tune the pollution process (i.e. self-enrichment) in planet hosts by not allowing
them to have [Fe/H] $>$ 0.5. Most probably the cutoff represents a rough upper limit to 
metallicities in the solar neibourhood. If [Fe/H]$\sim$0.4 represents the ``present day'' 
state of Galactic chemical evolution, then certain trends should appear for all other 
chemical species. How then can we  disentangle the abundance anomalies produced by 
the presence of planets ? 

The most simple way is to study those trends which are difficult to interpret in the 
framework of standard Galactic chemical evolution models. Given the constant rate of
Type II and Type Ia SN during the last 10 Gyrs of  Galactic evolution, we would
not expect any significant change in the slope of 
[$\alpha$/Fe] versus [Fe/H] where the $\alpha$'s are: O, Si, S, Mg, Ca and Ti. However, 
observations (Gonzalez et al.\,2001, Sadakane et al.\,2002, Bodaghee et al.\,2003) show a 
sudden change at [Fe/H] = 0 in the slopes of [Si/Fe] and [Ti/Fe] versus [Fe/H] while [Ca/Fe] 
decreases monotonically with [Fe/H]. 
It is not clear why Si and Ti should drastically change their slopes at [Fe/H]=0 and become flat. 
Moreover, high quality observations by Bensby et al.\,(2003) demonstrate that [O/Fe] 
continues to decrease at [Fe/H] $>$ 0 without showing the flatening out found in 
previous studies (Nissen \& Edvardsson\,1992). 
Galactic chemical evolution models predict similar trends for O, Si and all the other 
$\alpha$-elements (Tsujimoto et al.\,1995). Why C and N should have a different behaviour at high 
metallicities, both being volatiles and having similar production sites in the Galaxy ? 
There can be three reasons for these anomalies; a) models of Galactic chemical evolution
are very uncertain at high metallicities, b) abundance trends in metal-rich stars are 
affected by the presence of planets and c) abundance analysis of metal-rich stars is not reliable.

\begin{figure}
\plotfiddle{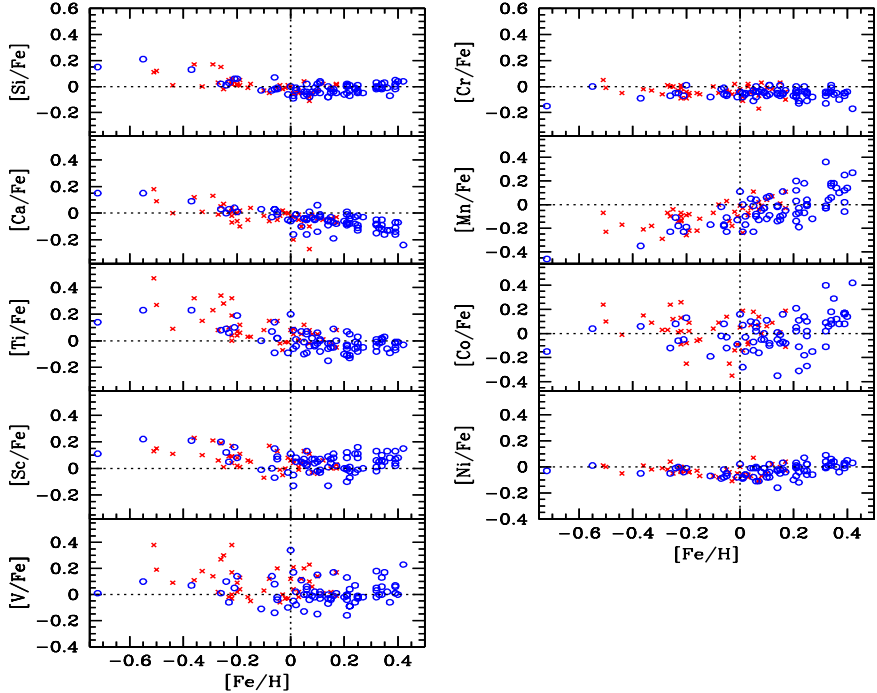}{8.5cm}{360}{60}{50}{-180}{-90}
\caption{[X/Fe] vs [Fe/H] plots for several $\alpha$- and Fe-group elements. The crosses
represent the comparison-sample stars, while the open circles denote the planet-hosting stars. 
From Bodaghee et al. (2003).}
\end{figure}
 
\subsection{Astrobiology}

Possible explanations for the abundance anomalies or the correlation between [Fe/H] and
the presence of giant planets have certain implications for astrobiology and even SETI. 
Already in the 1960's Drake (1965) and Shklovski (1966) independently proposed a method
which extraterrestrial intelligent civilizations could employ in order to announce 
their existence. They could add some short-lived isotope(s) into the atmosphere of their 
``Sun'' with the hope that possible observers would detect the absorption spectral lines of that
element and realize their
artificial origin. The amount of matter required to produce observable absorption lines 
of some rare elements/isotopes is not that large and any developed civilization should 
be able to handle this task. 

The possible correlation between [Fe/H] and the formation of giant planets may have
an impact on the formation of terrestrial planets and their habitability via a shielding effect.
Terrestrial planet formation should strongly depend on metallicity as well. 
A giant planet can relocate planetesimals and comets via scattering into the inner region of
the planetary system. This process can protect inner habitable terrestrial planets against the 
impacts of the comets. 
Numerical simulations (Ida, Junko \& Lin\,2001) show that a planet with large semi-major
axis and/or large mass may eject planetesimals and prevent pollution of metals onto the host star. 
These authors proposed that the habitability may be regulated by a giant planet(s)
since the shielding effect does not only inhibit impacts onto the host star but also prevents 
inner terrestrial planets from being impacted by cometary bodies. 

The Earth contains a generous amount of all the stable and quasi-stable elements present
in the Universe. The abundances of elements in the Sun and on the primitive Earth was
suited to the creation and evolution of plants and living beings. Even the rather rare elements, those
especially suited to the extraction of energy, using nuclear processes (like U and Th) are
present on the Earth in suitable concentration for the development of life. It is well known
that the rock masses that form the external part of the Earth have been in rapid 
motion for hundreds of millions of years. This motion is so rapid that $\twothirds$ of the
crust has been recycled into the Earth's mantle in the last 200 million years. 
Many geological processes (such as volcanic activity, drift of continents, iron catastrophe)
in the lithosphere result from the heat released by the radioactive elements, 
mostly Thorium and Uranium. 
Active volcanoes are a source of many volatile compounds (water, methane, carbon oxides, etc.)
which accumulated with time and formed the primitive gaseous atmosphere and, somewhat
later, the liquid hydrosphere. 

The above may suggest that there are certain chemical and physical preconditions for mankind's  
evolution. These preconditions (or ``initial conditions''), depend on the chemical composition 
and evolution of the protoplanetary matter. Looking back, we can now see how the products 
of supernova explosions many billion years ago have influenced the life on our planet. 
Abundance studies in stars with exoplanets may help us to understand the chemical evolution of the planetary
systems and learn about the evolution of life in the Universe.  

\acknowledgments
I would like to thank my colleagues Nuno Santos, Michel Mayor, Rafael Rebolo, Alexandra Ecuvillon,
Alister Graham and Ram\'on Garc\'\i a L\'opez for many useful discussions and their comments on the
text.

\end{document}